\documentstyle[epsf,aps]{revtex}
\begin{document}

\thispagestyle{empty}
{\baselineskip-4pt
\font\yitp=cmmib10 scaled\magstep2
\font\elevenmib=cmmib10 scaled\magstep1  \skewchar\elevenmib='177
\leftline{\baselineskip20pt\vbox to0pt
   { {\yitp\hbox{Osaka \hspace{1.5mm} University} }
     {\large\sl\hbox{{Theoretical Astrophysics}} }\vss}}

\rightline{\large\baselineskip20pt\rm\vbox to20pt{
\baselineskip14pt
\hbox{OU-TAP-35} 
\vspace{2mm}
\hbox{\today}\vss}}%
\vskip15mm
\begin{center}{\large \bf
Can the Simplest Two-Field Model of Open Inflation Survive? }\\
\vspace{0.7cm}
 {Misao Sasaki and Takahiro Tanaka 
 } \\
\vspace{0.3cm}
{\em Department of Earth and Space Science\\
Osaka University,~Toyonaka 560,~Japan}\\

\vspace{0.4cm}
\end{center}

\begin{abstract}
We investigate the quantum fluctuations of an
inflaton field in a two-field model of one-bubble open inflation.
One of the inflatons is assumed to be responsible for the first false
vacuum stage of inflation and the other is assumed to be
responsible for the second slow-roll stage of inflation.
The mass of the second inflaton is assumed to be negligible
throughout the whole era of inflation.
We find the super-curvature fluctuations are
 enhanced by the factor given 
by the ratio of the Hubble constants at false vacuum and 
true vacuum. This gives a strong constraint on a class of
open inflation models. In particular, this implies the simplest
two-field model proposed by Linde and Mezhlumian is in trouble.
\\ 

\noindent
{\normalsize \bf PACS number(s)}: 04.62.+v,98.80.Cq
\end{abstract}

\vspace{1cm}
There are growing observational evidences
that our universe has negative spatial curvature\cite{RPa}.
Accordingly an inflationary universe scenario which predicts 
an open but homogeneous and isotropic universe has been
proposed and studied\cite{BGT,YST95,Lindea,Lindeb}. 
We call it open inflation or one-bubble inflation. 

This scenario is composed of two inflationary phases 
separated by the bubble nucleation due to quantum tunneling. 
As a result of the first inflation, the universe becomes 
sufficiently homogeneous and isotropic. 
Then a vacuum bubble is formed by quantum tunneling. 
This process is called the false vacuum decay. 
It is dominated by the so-called $O(4)$-symmetric
bounce\cite{Col,DeLCol}. 
Reflecting this symmetry, the region inside 
the bubble can be regarded as an open Friedmann-Robertson-Walker
 universe. Subsequently the second slow-roll inflation
occurs inside the bubble and solves the entropy problem. 

Several models for inflaton potentials were 
proposed in this context\cite{BGT,Lindea,Lindeb}. 
Among them, the one that requires the least tuning of the model
 parameters is the simplest 
two-field model proposed by Linde and Mezhlumian\cite{Lindeb}. 
In this model, we consider two inflaton fields,
$\sigma$ and $\phi$, which have no interaction between them. 
The first inflaton field $\sigma$ has a tilted double 
well potential and is supposed to be
in the false vacuum initially. 
The vacuum energy of $\sigma$ dominates the universe 
and is responsible for the first stage of inflation. 
The second inflaton field $\phi$ has a simple 
quadratic potential $1/2\,m^2\phi^2$ with the mass $m$ 
being assumed to be small 
compared with the expansion rate of the universe $H(t)$,
just as in the chaotic inflation scenario.
After the bubble nucleation occurs, the potential energy of 
$\sigma$ disappears. 
Then the potential energy of $\phi$
dominates 
and the slow-roll inflation occurs in the bubble. 
It is important to note that $\dot\phi$ at the first stage of inflation
is much smaller than $\dot\phi$ at the second stage of inflation in this
scenario, to ensure the $O(4)$ $\bigl(O(3,1)\bigr)$ -symmetry of the
background. This means that the Hubble constant at the first stage of
inflation is much bigger than that at the second stage of inflation.

At the classical level, this model works well.
However, it is still an open question if it is consistent with 
the observed spectrum of density fluctuations of our universe.
In this letter, we estimate the amplitude of
curvature perturbations predicted by a class of two-field
open inflation models which includes the simplest model mentioned above. 

First we state several working hypotheses. 
In order to predict the spectrum of perturbations
of the universe, we need to specify their dominant origin. 
Here we assume that 1) the perturbations originate from 
the quantum fluctuations of the second inflaton field $\phi$,
which we denote by $\varphi(x)$,
 2) the mass of $\varphi$ is always negligible
compared with the Hubble parameter $H(t)$
 as in the simplest two-field model, and
3) $\phi$ has no strong coupling with other fields. 
Thus we approximate the fluctuation 
field $\varphi$ as a noninteracting massless scalar field. 
Since the first stage of inflation is supposed to 
last long enough, we also assume that 4) the quantum state 
of $\varphi$ is in the Euclidean (Bunch-Davies) vacuum before the 
false vacuum decay.
Furthermore we assume that 5) the bubble wall 
between true and false vacua can be described by using 
the thin wall approximation. However, we will see in the end
that relaxing the last assumption does not affect the essential 
part of the result.
Under these assumptions, we can use the formulae given in 
our previous paper\cite{paper1}(Paper I) 
to calculate
the amplitude of curvature perturbations on the comoving 
hypersurface, ${\cal R}_c$.

The classical description of the spacetime
is given by the $O(3,1)$-symmetric bubble.
In what follows, we denote this classical background field configuration
by $\phi_B$.
In the thin-wall approximation, the whole spacetime 
is given by the junction of two different de Sitter 
spaces with different Hubble constants $H_L$ and $H_R$. 
We take the region with the Hubble constant $H_L$ to be the false
vacuum. Hence $H_L > H_R$.
We divide the spacetime into six regions as shown in Fig.~1. 
The coordinates in the regions $R$, $L$, $C_R$ and $C_L$ 
are given by 
\begin{eqnarray}
 ds_{J}^2 & = & 
  H_J^{-2}\left[ -dt_J^2+\sinh^2 t_J \left
  (dr_J^2+\sinh^2 r_J d\Omega^2\right)\right],
\cr
  &&\hspace{2cm} \hbox{in $R$ and $L$},
\cr
 ds_{C,J}^2 & = & 
  H_J^{-2} \left[dt_{C,J}^2+\cos^2 t_{C,J}
  \left(-dr_{C}^2+\cosh^2 r_{C} d\Omega^2\right)\right],
\hspace{-2cm}
\cr
  &&\hspace{2cm}  \hbox{in $C_R$ and $C_L$},
\nonumber \\
\end{eqnarray}
where $J=R$ or $L$.
The relations among the coordinate systems are uniquely determined 
by the analyticity which was discussed in Paper I. 
Here we write the resulting relations, 
\begin{eqnarray}
 t_R=i t_{C,R}-\pi i/2,\quad r_R=r_{C}+\pi i/2,
\cr
 t_L=-i t_{C,L}-\pi i/2,\quad r_L=r_{C}+\pi i/2.
\label{CoTr}
\end{eqnarray}
For later convenience, we introduce the coordinate $T$ which 
covers both regions $C_R$ and $C_L$. Using $T$, the metric there
is written as 
\begin{equation}
 ds_{C}^2 = 
  dT^2+a^2(T)
  \left(-dr_{C}^2+\cosh^2 r_{C} d\Omega^2\right),
\end{equation}
where 
\begin{equation}
    dT={dt_{C,J}\over H_J}\raisebox{-5pt}{,}
    \quad a(T)={\cos t_{C,J} \over H_J}\raisebox{-5pt}{,}
\end{equation}
in the region $C_J$.
The junction condition of the metric 
requires the continuity of the scale factor $a(T)$ on the wall,
\begin{equation}
{\cos t_{C,L}\over H_L}\Bigr\vert_{wall}
={\cos t_{C,R}\over H_R}\Bigr\vert_{wall}\,\raisebox{-5pt}{.}
\end{equation}

In order to calculate ${\cal R}_c$, we need to solve the field equation 
for $\varphi$.
First we consider the region $C$. 
By using the harmonics, 
\begin{equation}
Y_{p\ell m}={\Gamma(ip+\ell+1)\over\Gamma(ip+1)}
   {p\over\sqrt{\sinh r_J}}P^{-\ell-1/2}_{ip-1/2}
   (\cosh r_J)Y_{\ell m}(\Omega),
\end{equation}
and its analytic extension to the region $C$, 
we write the solution in the decomposed form as
\begin{equation}
 u_{p\ell m}(x)={u_{p}(T)}Y_{p\ell m}(r_C,\Omega).
\end{equation}
Then the field equation in region $C$ becomes 
\begin{equation}
 \left[{1\over a^3(T)}{\partial\over \partial T} a^3(T)
   {\partial\over \partial T} + {p^2+1\over a^2(T)}\right] 
   {u_{p}(T)}=0.
\end{equation}
The field equations in the other regions are obtained by the 
analytic continuation. 

As stressed in our previous paper\cite{STY95}, 
the perturbation modes are divided into two classes. 
One has usual continuous spectrum with $p^2>0$ and 
the other has discrete spectrum with $p^2<0$. 
We call the latter super-curvature modes since their characteristic
wavelengths are greater than the spatial curvature scale. 

We first consider the continuous modes.
We introduce the fundamental mode functions which are most naturally
defined in regions $R$ and $L$;
\begin{eqnarray}
u_{p}^{(R)} & = & {H_R\over \sinh t_R}{\cosh t_R-ip\over \Gamma(2-ip)}
       \left({\cosh t_R+1\over \cosh t_R-1}\right)^{ip/2}        
\nonumber \\
 & = & {H_R e^{-\pi p/2}\over -i\cos t_{C,R}}
      {\sin t_{C,R} -ip\over \Gamma(2-ip)}
       \left({1+\sin t_{C,R}\over 1-\sin t_{C,R}} \right)^{ip/2}
   \hspace{-5mm}\raisebox{-5pt}{,}
\end{eqnarray}
and
\begin{eqnarray}
u_p^{(L)} & = & {H_L\over \sinh t_L}
       {\cosh t_L-ip\over \Gamma(2-ip)}
       \left(\cosh t_L+1\over \cosh t_L-1\right)^{ip/2}
\nonumber \\
 & = & {H_L e^{-\pi p/2}\over -i\cos t_{C,L}}
      {\sin t_{C,L} +ip\over \Gamma(2-ip)}
       \left(1+\sin t_{C,L}\over 1-\sin t_{C,L}\ \right)^{-ip/2}
  \hspace{-7mm}\raisebox{-5pt}{.}
\end{eqnarray}
In the second line of each equation, 
the functions are naturally extended to the region $C_J$ by the 
analytic continuation.
The extension of these functions 
to the other side of the wall is given by matching
the mode functions at the wall. 
Since the first derivative of the scale factor, $\dot a(T)$, 
where the dot means the derivative with respect to $T$, 
is discontinuous but finite at the wall, 
the field equation requires the continuity of 
$u_{p}$ and $\dot u_{p}$. Putting
\begin{equation}
u_{p\ell m}:= u_{p\ell m}^{(R)}=\alpha_p u_{-p\ell m}^{(L)}
      +\beta_p u_{p\ell m}^{(L)},
\end{equation}
the junction condition determines $\alpha_p$ and $\beta_p$ as
\begin{eqnarray}
\alpha_p &=&((u_p^{(L)},u_{p}^{(R)}))
   /((u_p^{(L)},u_{-p}^{(L)})),\cr
\beta_p &=&((u_p^{(R)},u_{-p}^{(L)}))
   /((u_p^{(L)},u_{-p}^{(L)})),
\label{alpbet}
\end{eqnarray}
where
\begin{equation}
((u,v))={1\over a(T)}\left(u\dot v-\dot u v\right)\biggr\vert_{wall}
  \raisebox{-5pt}{.}
\end{equation}
Then one can construct the orthonormalized mode functions
$v_{p\sigma\ell m}$ ($\sigma=\pm1$) over the whole spacetime
by forming linear combinations of
$u_{p\ell m}$ and $u_{-p\ell m}$ (Paper I), which we write
as $v_{p\sigma\ell m}=V_{p\sigma}Y_{plm}$.

Next we turn to the discrete mode. By setting $p=i\Lambda$, 
we write the fundamental mode function as
\begin{equation}
 u_{\Lambda\ell m}^{(J)}=u_{\Lambda}^{(J)} \tilde Y_{\Lambda \ell m},
\end{equation}
where
\begin{equation}
 \tilde Y_{\Lambda \ell m}={P^{-l -1/2}_{-\Lambda -1/2}(\cosh r)\over
  \sqrt{\sinh r}} Y_{\ell m}(\Omega),
\end{equation}
and
\begin{eqnarray}
u_{\Lambda}^{(R)} & = & {H_R\over \sinh t_R}
       {\cosh t_R+\Lambda\over\Gamma(2+\Lambda)} 
       \left(\cosh t_R+1\over \cosh t_R-1\right)^{-\Lambda/2}
\nonumber \\
 & = & {H_Re^{-i\pi \Lambda/2}\over -i\cos t_{C,R}}
       {\sin t_{C,R} +\Lambda\over\Gamma(2+\Lambda)} 
     \left(1+\sin t_{C,R}\over 1-\sin t_{C,R}\ \right)^{-\Lambda/2}
  \hspace{-7mm}\raisebox{-5pt}{,}
\nonumber \\
u_{\Lambda}^{(L)} & = & {H_L\over \sinh t_L}
       {\cosh t_L+\Lambda\over\Gamma(2+\Lambda)} 
       \left(\cosh t_L+1\over \cosh t_L-1\right)^{-\Lambda/2}
\nonumber \\
 & = & {H_Le^{-i\pi \Lambda/2}\over -i\cos t_{C,L}}
       {\sin t_{C,L} -\Lambda\over\Gamma(2+\Lambda)}
      \left(1+\sin t_{C,L}\over 1-\sin t_{C,L}\ \right)^{\Lambda/2}
  \hspace{-5mm}\raisebox{-5pt}{.}
\end{eqnarray}
Note that 
the regularity condition at $t_J=0$ determines
the form of $u_\Lambda^{(J)}$ as given above, and
further requires $\Lambda>0$.
Then the continuity condition of the mode function gives
\begin{equation}
 {1-\Lambda^2\over\sin t_{C,R} +\Lambda}
 ={1-\Lambda^2\over\sin t_{C,L} -\Lambda}
  \raisebox{-5pt}{,} 
\end{equation}
which determines the eigenvalue $\Lambda$.
One easily sees the only positive root is $\Lambda=1$. 
The corresponding mode also exists in the case of the pure 
de Sitter background\cite{STY95}. 
Hence we call it the de Sitter super-curvature mode. 
In the present case, the explicit form of the mode function becomes 
\begin{equation}
 u_\Lambda=u_\Lambda^{(R)}= {H_R\over 2}
  \raisebox{-5pt}{,}
\end{equation}
in the entire region of $C$.

To determine the amplitude of the curvature perturbation due to 
this mode, we need to calculate the normalization factor
$N_{\Lambda}$, which is defined by
\begin{eqnarray}
 N_{\Lambda} & := & \int dT\,a(T)\vert u_{\Lambda}(T)\vert^2
\cr
   & = & {1\over H_R^2} \int^{\pi/2}_{wall} dt_{C,R} \cos t_{C,R}
      \vert u_{\Lambda}\vert^2
\cr &&    
   +{1\over H_L^2} \int_{-\pi/2}^{wall} dt_{C,L} \cos t_{C,L}
      \vert u_{\Lambda}\vert^2.
\label{norm}
\end{eqnarray}
In the present case, $N_{\Lambda}$ is evaluated to be
\begin{equation}
 N_{\Lambda} =2 \left({H_R\over H_L}\right)^2
    \left(1+s_L\over 1+s_R\right)\left(1+{\Delta s\over 2}\right)
   \raisebox{-5pt}{.}     
\end{equation}
The normalized mode function is then given by
$V_\Lambda=N_\Lambda^{-1/2}u_\Lambda$.

Now let us evaluate the curvature perturbation on the comoving
hypersurface. We expand ${\cal R}_c$ by modes,
\begin{equation}
 {\cal R}_c =\sum_{\ell m}\int_0^{\infty}
  dp~{\cal R}_{p} Y_{p\ell m}(r,\Omega)
  +\sum_{n,\ell,m}{\cal R}_{\Lambda_n}
\tilde Y_{\Lambda_n\ell m}(r,\Omega).
\end{equation}
We first consider the contribution from the continuous modes. 
The power spectrum of the curvature perturbation ${\cal R}_p$
is given by
\begin{equation}
  \vert {\cal R}_p \vert^2=\left({H_R\over\dot\phi_B}\right)^2
       \lim_{t_R\to\infty}\sum_{\sigma=\pm1}|V_{p\sigma}|^2,
\end{equation}
which is expressed as
\begin{equation}
|{\cal R}_p|^2=|{\cal R}_p|_{BD}^2(1-Y),
\end{equation}
where
\begin{equation}
 \vert {\cal R}_p \vert^2_{BD}
   \left({H_R^2\over \dot\phi_B}\right)^2
  {\coth \pi p\over 2p(1+p^2)}
  \raisebox{-5pt}{,}
\end{equation}
is the spectrum for the Bunch-Davies vacuum, and 
$Y$ is expressed in terms of $\alpha_p$ and $\beta_p$ as 
\begin{equation}
  Y={\Gamma(2-ip)\over \Gamma(2+ip)}
  {e^{-\pi p}\beta_p\over 2\cosh \pi p~\bar \alpha_p}+\hbox{c.c.}\,.  
\end{equation}
From the Eq.~(\ref{alpbet}), 
after a straightforward calculation, we obtain 
\begin{eqnarray}
 \alpha_p & = & e^{-\pi p}{\Gamma(2+ip)\over\Gamma(2-ip)}
           \left({(1+s_R)(1-s_L)\over (1-s_R)(1+s_L)}\right)^{ip/2}
           \left(1+{i\Delta s\over 2p}\right)
   \raisebox{-5pt}{,}\hspace{-2mm}
\nonumber \\
 \beta_p & = & -\left({(1+s_R)(1+s_L)\over (1-s_R)(1-s_L)}\right)^{ip/2}
           {i\Delta s\over 2p}\raisebox{-5pt}{,}
\end{eqnarray}
where 
$s_R:=\sin t_{C,R}\vert_{wall}$, $s_L:=\sin t_{C,L}\vert_{wall}$ 
and $\Delta s:=s_R -s_L$. Note that $(1-s_R^2)/H_R^2=(1-s_L^2)/H_L^2$.
Thus we finally obtain
\begin{equation}
1-Y =
 1-{(\Delta s)^2 \cos\tilde p
     +2p \Delta s\sin \tilde p
      \over \cosh \pi p\,(4p^2+(\Delta s)^2)}\raisebox{-5pt}{,}
\end{equation}
where
\begin{equation}
 \tilde p=p\ln\left(\displaystyle{1+s_R\over 1-s_R}\right)
 \raisebox{-5pt}{.}
\end{equation}
When $p$ is large, the second term in the right hand side 
is exponentially suppressed due to the factor, $1/\cosh\pi p$. 
On the other hand, when $p$ is small, it goes to $1$. 
Thus it is expected that $1-Y$ does not become much larger than unity.
Therefore we conclude that the fluctuations due to the continuous 
modes are never enhanced much in comparison with 
those in the Bunch-Davies vacuum. That is, we have
$\varphi\sim H_R$ inside the bubble although $\varphi\sim H_L$ outside
the bubble,  which is in accordance with an intuitive argument
given by Linde and Mezhlumian\cite{Lindeb}.

As for the contribution from the de Sitter super-curvature mode,
we apply the formula obtained in Paper I,
\begin{equation}
|{\cal R}_\Lambda|^2=\left({H_R\over\dot\phi_B}\right)^2|V_\Lambda|^2\,.
\end{equation}
Then 
\begin{equation}
 \vert {\cal R}_{\Lambda}\vert^2=
   {2\over N_{\Lambda}} \vert {\cal R}_{\Lambda}\vert^2_{BD},
\end{equation}
where 
\begin{equation}
  \vert {\cal R}_{\Lambda}\vert^2_{BD}={1\over 2}\left({
    H_R^2\over \dot \phi_B}\right)^2
  \raisebox{-5pt}{,} 
\end{equation}
is the amplitude in the Bunch-Davies vacuum limit. 
Thus we conclude 
\begin{equation}
 \vert {\cal R}_{\Lambda}\vert^2= \vert {\cal R}_{\Lambda}\vert^2_{BD}
   \times O\left({H_L^2\over H_R^2}\right)\raisebox{-5pt}{.}
\label{amp}
\end{equation} 
This means that the suppression mechanism as discussed by
Linde and Mezhlumian\cite{Lindeb} does not work for
the de Sitter super-curvature mode, contrary to the case 
for the continuous modes.
This difference is caused by the fact 
that the normalization of the
mode functions for the continuous modes is determined 
essentially by their behaviors at $t_{C,R}\to\pi/2$ and
$t_{C,L}\to-\pi/2$\cite{STY95},
while that for the super-curvature mode is determined by 
its behavior over the whole region $C$.

In the above we have used the thin-wall approximation.
However the conclusion will not change even if the wall is thick.
In order to understand this, we note that
$u_\Lambda=const.$ is a solution of the field equation with 
$p=-i$ independent of the detail of $a(T)$.
Hence we have $N_{\Lambda}=\vert u_{\Lambda}\vert^2\int dT\,a(T)$.
Since this integral is determined predominantly by the Hubble radius
$H_L$ of the false vacuum region, one has $\int dT\,a(T)\sim H_L^{-2}$
Therefore $N_{\Lambda}=O(H_R^2/H_L^2)$ for $u_\Lambda=H_R/2$. 
Thus we conclude that Eq.~(\ref{amp}) holds generally 
independent of the thin-wall approximation.

Now let us consider implications of our results.
We have found that the amplitude of the perturbations
due to the continuous modes are not much different from
the case of the Bunch-Davies vacuum but that due to
the de Sitter super-curvature mode is enhanced by a factor
of $O(H_L/H_R)$.
In one of our previous papers\cite{YST95}, we 
discussed the effect of the super-curvature mode on the 
power spectrum of the cosmic microwave background (CMB) anisotropies
 $C_{\ell}$.
There we assumed the Bunch-Davies vacuum. We found
the contribution from the super-curvature mode reaches more than
 $10\%$ for models with $\Omega_0<0.2$. 
On the other hand, it gives only a negligible contribution to 
the CMB spectrum at $\ell>10$. 
Since only the super-curvature mode are enhanced 
in models we have studied in this letter, 
the models will predict a large enhancement of 
the CMB spectrum at $\ell<10$ if $H_L\gg H_R$,
 which will be in contradiction with observed spectrum\cite{YB}.
Thus the simplest two field model proposed by Linde and
Mezhlumian\cite{Lindeb} is in trouble, though the other two-field
models proposed by them are still viable.

In conclusion, in the context of one-bubble open inflation,
we have investigated the quantum fluctuations of an
inflaton field whose mass is negligible at both the first and second
stages of inflation and which is responsible for the second
stage of inflation.
We have found the super-curvature fluctuations are
 enhanced by the factor given
by the ratio of the Hubble constants at false vacuum and 
true vacuum.
This gives a strong constraint on a class of open inflation models
in which the Hubble constant outside a vacuum bubble
 is much greater than that inside the bubble. 

\vspace{5mm}
\begin{center}
{\bf ACKNOWLEDGMENTS}
\end{center}
We would like to thank A. D. Linde for calling our attention 
to this problem and K. Yamamoto for helpful comments. 

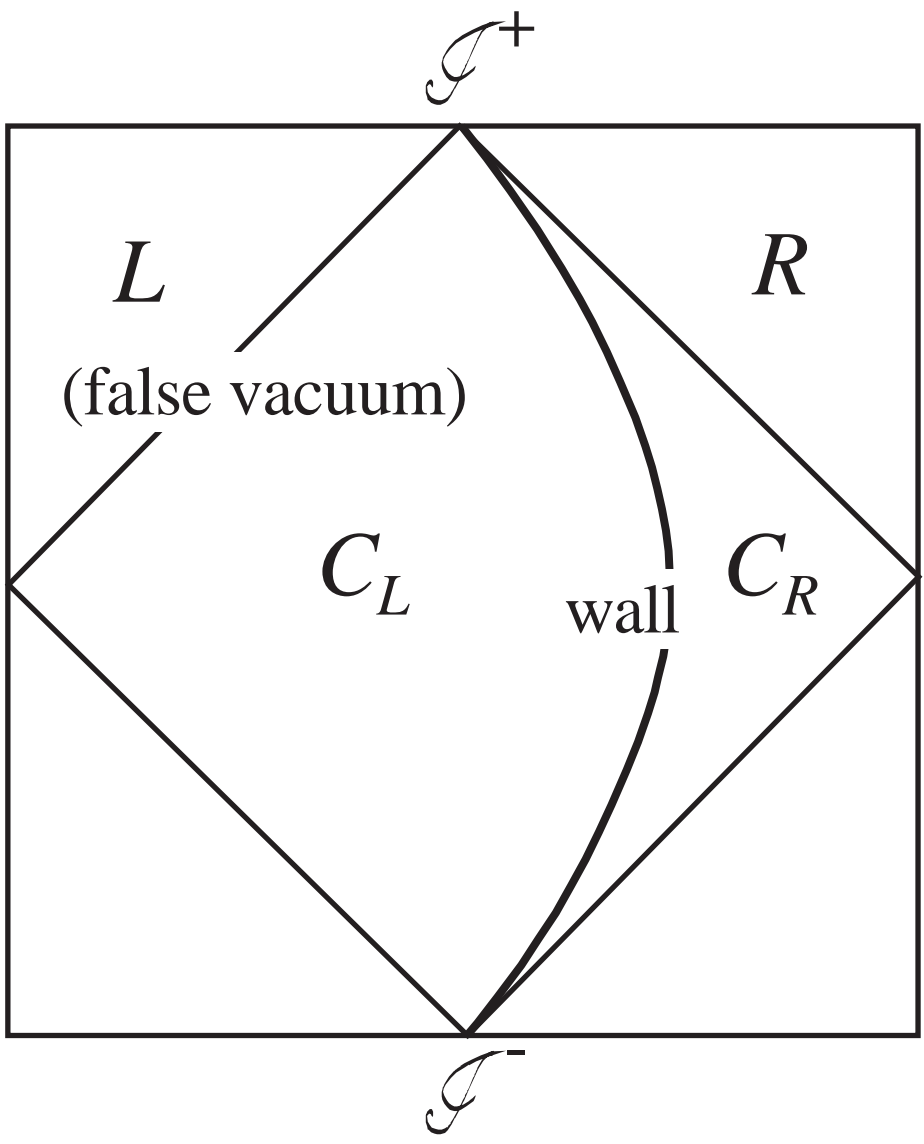
\end{document}